\documentclass[twocolumn,prl,showpacs,preprintnumbers,amsmath]{revtex4}
\usepackage{graphicx}
\usepackage{dcolumn}
\usepackage{bm}

\begin{document}

\title{Topological Change of the Fermi Surface in Low Density Rashba 
Gases: Application to Superconductivity}
\author{E. Cappelluti$^{1,2}$}

\author{C. Grimaldi$^{3,4}$}

\author{F. Marsiglio$^{4,5}$}

\affiliation{$^1$SMC-INFM and Istituto dei Sistemi Complessi, CNR-INFM, via dei
Taurini 19, 00185 Roma, Italy}

\affiliation{$^2$Dipart. di Fisica, Universit\`a ``La Sapienza'',
P.le A. Moro 2, 00185 Roma, Italy}

\affiliation{$^3$ LPM, Ecole Polytechnique F\'ed\'erale de
Lausanne, Station 17, CH-1015 Lausanne, Switzerland}

\affiliation{$^4$DPMC, Universit\'e de Gen\`{e}ve, 24 Quai
Ernest-Ansermet, CH-1211 Gen\`{e}ve 4, Switzerland}

\affiliation{$^5$Department of Physics, University of Alberta,
Edmonton, Alberta, Canada, T6G 2J1}

\begin{abstract}
Strong spin-orbit coupling can have a profound effect on the
electronic structure in a metal or semiconductor, particularly for
low electron concentrations. We show how, for small values of the 
Fermi energy compared to the spin-orbit splitting of Rashba type, 
a topological change of the Fermi surface leads to an effective 
reduction of the dimensionality in the electronic 
density of states. We investigate its consequences on the
onset of the superconducting instability. We show,
by solving the Eliashberg equations 
for the critical temperature as a function of spin-orbit 
coupling and electron density, that the superconducting
critical temperature is significantly tuned in this regime
by the spin-orbit coupling. We suggest that materials
with strong spin-orbit coupling are good candidates for enhanced 
superconductivity.
\end{abstract}

\maketitle

Spin-orbit (SO) coupling arising from the
lack of inversion symmetry plays a leading role
in the field of spintronics \cite{wolf01}. One of the main goals
in this field of research is the possibility of
tuning the electron spin properties (transport, coherence, relaxation,
etc.) by means of electrical fields \cite{sharma05}.
With this aim in mind, different properties have been investigated,
such as spin relaxation \cite{spin_relaxation},
magnetoconductance \cite{magnetoconductance} and spin-Hall currents 
\cite{engel06}.
As a general rule, the spin-orbit coupling is assumed to be
quite small
with respect to the other relevant energy scales,
in particular with respect to the electronic dispersion, so that
the infinite bandwidth limit is often employed.
While this assumption is indeed rather reasonable in most of the cases,
the natural aim of the current investigations is
to search for new materials with stronger spin-orbit couplings,
as for instance in HgTe quantum wells \cite{gui04}, or
the surface states of metals and semimetals \cite{rotenberg99,koroteev04}.
For this reason, experimental evidence of a Rashba SO coupling with
energy $E_0$  (to be defined below) as large as $\simeq 220$ meV in 
bismuth/silver alloys \cite{ast06},
or with $E_0 \simeq 30-200$ meV in non-centrosymmetric
superconductors CePt$_3$Si \cite{bauer04,samokhin04}, Li$_2$Pd$_3$B, and 
Li$_2$Pt$_3$B \cite{togano04,yuan06},
is certainly an important step towards the investigation of new materials
with large SO coupling.
Needless to say, the existence of such materials compels us to carry 
out a more thorough investigation of the properties of SO systems
when $E_0$ is no longer necessarily the smallest energy scale
in the problem.

The possibility of having novel interesting features
in low density systems with
Fermi energy $E_{\rm F}$ of the same order or lower than
the SO energy $E_0$, in particular,
has not been sufficiently investigated to date, in
our opinion, and only few studies have been devoted to this problem.
In Ref. \cite{grimaldi06a} for instance,
it was shown that the vanishing of the spin-Hall current
in the low density limit $n \rightarrow 0$ of Rashba disordered systems
is not related to the vanishing
of the vertex function (which applies only
in the strict $E_{\rm F}/E_0 \rightarrow \infty$ limit)
but rather to the cancellation between on-Fermi surface
and off-Fermi surface contributions.
Another interesting effect was also
pointed out in Ref. \cite{grimaldi05}: there
the spin relaxation time $\tau_s$
for $E_{\rm F} \ll E_0$
was shown to be proportional to the
electron scattering time $\tau$, in contrast with
the standard Dyakonov-Perel behavior, where
$\tau_s$ scales as $1/\tau$ \cite{dyakonov71}.
On the other hand, in spite of the growing interest in the
properties of non-centrosymmetric superconductors
with strong SO coupling \cite{gorkov01,frigeri04}, no specific
investigation to explore the regime where
$E_{\rm F}/E_0 \lesssim 1$ has been pursued, to our knowledge.

The aim of this Letter is to explore in detail a fundamental feature
arising in SO Rashba systems,
namely the topological change of the Fermi surface
induced by the strong SO interaction in the low density regime
$E_{\rm F}/E_0 \le 1$. We show that in this situation
the enhanced phase space available for the electronic excitations
gives rise to a SO-induced change of the
electronic density of states which can be described in terms
of an effective reduced dimensionality.
We discuss the consequences of this scenario
on the superconducting instability criterion for both
two- and three-dimensional Rashba systems.
We show that, in contrast with the high density case
$E_{\rm F}/E_0 \gg 1$,
the SO coupling in the $E_{\rm F}/E_0 \lesssim 1$ regime
systematically enhances the superconducting critical temperature $T_c$,
providing evidence that the lack of inversion
symmetry can be remarkably beneficial for superconducting pairing.

We begin our analysis by considering the Rashba
model \cite{rashba60} which describes the linear coupling 
of conduction electrons with a spin-orbit potential of the form
\begin{equation}
\gamma (k_x\sigma_y-k_y\sigma_x),
\label{SO}
\end{equation}
where $\sigma_x$, $\sigma_y$
are Pauli matrices and $\gamma$ is the Rashba coupling constant.
For two-dimensional systems such as asymmetric quantum wells and
surface states of metals and semimetals
the SO coupling in Eq. (\ref{SO})
arises from the asymmetric confining potential, while
in bulk three-dimensional compounds
Eq. (\ref{SO}) originates from the lack of reflection
symmetry with respect to the $z$-direction,
as in CePt$_3$Si.
The SO coupling is reflected in an energy splitting of the two helicity
bands. Assuming a parabolic band for $\gamma=0$,
the resulting dispersion of the electronic excitations,
for two-dimensional (2D) and three-dimensional (3D) cases,
reduces to
\begin{eqnarray}
E^{\rm 2D}_\pm (k)
&=& \frac{\hbar^2}{2m^*}\left(k \pm k_0\right)^2,
\label{2d}
\\
E^{\rm 3D}_\pm (k) &=& \frac{\hbar^2}{2m^*}\left(k \pm k_0\right)^2
+\frac{\hbar^2 k_z^2}{2m^*},
\label{3d}
\end{eqnarray}
where $k=|{\bf k}|$ is the modulus of the $xy$ in-plane momentum
${\bf k}=(k_x,k_y)$, $k_z$ is the wavevector along the $z$-direction,
and $m^*$ is the effective electron mass.
In addition $k_0=m^*\gamma/\hbar^2$ represents here the characteristic
Rashba momentum.
\begin{figure}
\includegraphics[width=0.42\textwidth]{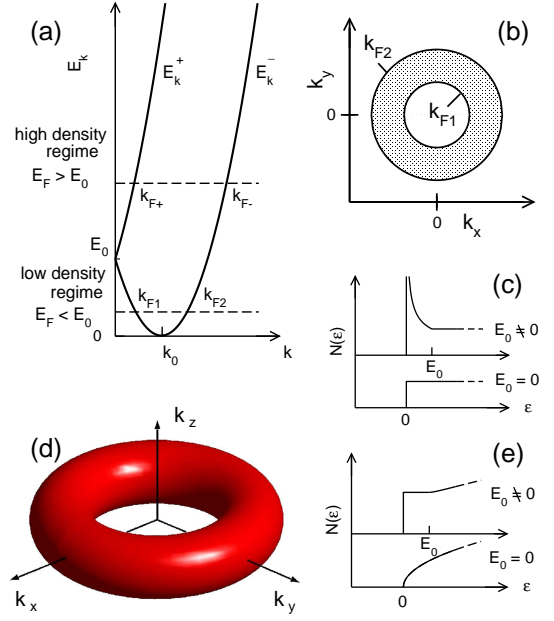}
\caption{Panel (a): electronic dispersion in the presence of SO
coupling: for $E_{\rm F} \ge E_0$ (high density regime) the two
Fermi surfaces belong to different helicity bands, while for
$E_{\rm F} \le E_0$ (low density regime) the Fermi surface exists
only on the $E_{\bf k}^-$ band. Also shown here is the Rashba
energy $E_0$ corresponding to the lowest interband excitation
energy for an electron lying at the bottom of the lower band.
Panels (b) and (c): Fermi surface and density of states,
respectively, in the low density regime for the two-dimensional
case. Panels (d) and (e): Fermi surface and density of states in
the low density regime for the three-dimensional case.}
\label{f-sketch}
\end{figure}
The dispersion for the 2D case is shown in Fig.~\ref{f-sketch}a, which can
easily be generalized in the 3D case by taking into account the
$k_z$ dispersion. The two horizontal dashed lines correspond
to the Fermi level for high density and low density regimes
defined as $E_{\rm F}> E_0$ and $E_{\rm F}< E_0$ respectively,
where $E_0=\hbar^2 k_0^2/2m^*$ is
the energy of the ${\bf k}=0$ point with respect
to the bottom band edge at $k=k_0$
(see Fig.~\ref{f-sketch}a).

{\em Density of states} -
Several studies in the literature have focused
on the high density regime, $E_{\rm F}\gg E_0$,
where the two Fermi surfaces belong
to different helicity bands, and where the
Fermi volume $V_{\rm F}$ is given by the area of two concentric Fermi
circles $V_{\rm F}=\pi k_{{\rm F},+}^2 + \pi k_{{\rm F},-}^2$
with $k_{{\rm F},\pm}=\sqrt{2m^*/\hbar^2}
\left( \sqrt{E_{\rm F}}\mp \sqrt{E_0}\right)$.
In this case the electronic density of states (DOS)
at the Fermi level is given by
\begin{equation}
N^{\rm 2D}(E_{\rm F})
=\sum_{s}
\frac{1}{4\pi^2}\int_{S_{{\rm F},s}}
\frac{dS_k}{\hbar |v_{k,s}|} =
\sum_{s}
\frac{1}{4\pi^2\hbar}
\frac{S_{{\rm F},s}}{| v_{{\rm F},s}|},
\label{dos2d}
\end{equation}
where the Fermi velocity
$|v_{{\rm F},\pm}|= \sqrt{2E_{\rm F}/m^*}$ is independent of
the helicity number $s=\pm$ and the Fermi surfaces are
$S_{{\rm F},\pm}= 2\pi k_{{\rm F},\pm}$.
Hence the total DOS in the $E_{\rm F}> E_0$ regime
$N^{\rm 2D}(E_{\rm F})=m^*/(\pi \hbar^2)$
is identical to the one in the absence
of spin-orbit coupling.
A similar result applies for the 3D case where,
from Eq. (\ref{3d}), the corresponding
DOS can be obtained as
\begin{equation}
N^{\rm 3D}(E_{\rm F})=\int \frac{dk_z}{2\pi}
N^{\rm 2D}(E_{\rm F}-\hbar^2k_z^2/2m^*).
\label{dos3d}
\end{equation}
In the high density regime $E_{\rm F}\ge E_0$
we get $N^{\rm 3D}(E_{\rm F})=a\{\sqrt{E_{\rm F}-E_0}
+\sqrt{E_0}\arctan[\sqrt{E_0/(E_{\rm F}-E_0)}]\}$,
where $a=\sqrt{2}{m^*}^{3/2}/(\pi^2\hbar^3)$,
which reduces to
$N^{\rm 3D}(E_{\rm F})\simeq a\sqrt{E_{\rm F}}$
in the $E_{\rm F}/E_0 \gg 1$ limit. This again is the result
one obtains in the absence of SO coupling.

Let us now consider the $E_{\rm F} \le E_0$ regime.
In this case the Fermi level intersects only the lower $E_-(k)$ band
and the topology of the Fermi surfaces drastically changes.
In the 2D case for instance only the annulus that lies
between two Fermi circles of radii
$k_{{\rm F},2}=\sqrt{2m^*/\hbar^2}
\left( \sqrt{E_0}+ \sqrt{E_{\rm F}}\right)$
and $k_{{\rm F},1}=\sqrt{2m^*/\hbar^2}
\left( \sqrt{E_0}- \sqrt{E_{\rm F}}\right)$,
belonging to the {\em same}
helicity band, is filled (Fig.~\ref{f-sketch}b), and the inner Fermi surface
is {\em inwards} oriented.
We can still employ Eq. (\ref{dos2d}) by summing over
the two Fermi surface indexes $s=1,2$.
Using $|v_{{\rm F},s}|= \sqrt{2E_{\rm F}/m^*}$
and $S_{{\rm F},s}= 2\pi k_{{\rm F},s}$ we get
\begin{equation}
N^{\rm 2D}(E_{\rm F})=
\frac{m^*}{\pi\hbar^2}\sqrt{\frac{E_0}{E_{\rm F}}},
\label{dos2dSO}
\end{equation}
which is valid as long as  $E_{\rm F} \le E_0$ (Fig.~\ref{f-sketch}c).
Most peculiar is the square-root divergence for
$E_{\rm F}\rightarrow 0$ that is reminiscent of
one-dimensional behavior.
We relate such a feature to the non-vanishing
in the low density limit of the Fermi surface
which remains finite, $S_{{\rm F},s} \propto \sqrt{E_0}$, while
at the same time, the Fermi velocity vanishes as $\sqrt{E_{\rm F}}$. 
The behavior of Eq. (\ref{dos2dSO}) has to be compared
with the $\gamma=0$ case where also the Fermi surface
shrinks as $\sqrt{E_{\rm F}}$
and the electron DOS has a non divergent step-like behavior in the
$E_{\rm F} \rightarrow 0$ limit \cite{note}. 

A similar reduction of effective dimensionality in the electron DOS
appears also for the 3D systems in the $E_{\rm F} \le E_0$ regime.
In this case the Fermi surface
has a torus-like topology as shown in Fig. \ref{f-sketch}d,
with major radius $k_0=\sqrt{2m^*E_0/\hbar^2}$ and
minor radius $\sqrt{2m^* E_{\rm F}/\hbar^2}$.
Applying once more Eq. (\ref{dos3d}) we get
\begin{equation}
N^{\rm 3D}(E_{\rm F})=\frac{\pi a}{2}\sqrt{E_0},
\label{dos3dSO}
\end{equation}
for $E_{\rm F} \le E_0$.
We see therefore that the SO coupling changes qualitatively
the low density behavior of the DOS providing a
finite step-like behavior (Fig. \ref{f-sketch}e) in contrast with
a standard 3D electron gas whose DOS vanishes as
$\sqrt{E_{\rm F}}$.

{\em Cooper instability} -
The above described reduction of effective dimensionality
sheds a new light on the possible existence of
a superconducting phase in the low density regime of
systems with no inversion symmetry.
To illustrate this point
let us consider the classical problem \cite{cooper56} of a Fermi surface
instability towards the formation of a Cooper pair:
\begin{equation}
1=V \int_0^{\omega_0}d\xi N(\xi) \frac{1}{2\xi+\Delta},
\label{cooper}
\end{equation}
where $\Delta > 0$ is the binding energy
of the bound pair state, $V$ is the strength of the
interaction which we consider here for simplicity in the $s$-wave channel,
and where we have introduced a standard BCS cut-off $\omega_0$
related to the energy of the underlying bosonic mediator.
Since the superconducting Cooper pairing is essentially
a Fermi surface instability,
the strength of the bound state and its existence itself
is intimately related to the
phase space of the available electronic excitations.
For instance,
as well known, in the low density limit of 3D systems,
where $N^{\rm 3D}(\xi) = a \sqrt{\xi}$,
Eq. (\ref{cooper}) predicts a finite critical
coupling $V_c=1/(a\sqrt{\omega_0})$ below which no
bound state exists.

This result changes drastically for finite SO couplings
where, as seen above, the electron DOS behaves now
as an effective 2D system with a constant value in the low energy regime.
Using Eq. (\ref{dos3dSO}) in Eq. (\ref{cooper}) we get:
\begin{equation}
\Delta_{\rm 3D}
\simeq  2\omega_0 \exp\left(-\frac{4}{\pi a V \sqrt{E_0}}\right),
\label{gap3dSO}
\end{equation}
which explicitly shows that, contrary to the usual
3D case, the Cooper pair instability exists no matter how weak $V$ is.
Furthermore Eq. (\ref{gap3dSO}) predicts that
the pair energy $\Delta$ has an exponential dependence on the SO coupling,
as is normally the case (here as well) for the attractive potential.

A similar change of the character of the Cooper instability
occurs also in the 2D case.
Indeed, in the absence of SO coupling, one would get
the standard BCS result
$\Delta=2\omega_0 \exp(-2\pi\hbar^2/m^* V)$.
On the other hand, due to the
strong one-dimensional-like divergence of the electron DOS,
Eq. (\ref{dos2dSO}), the binding energy
for finite SO coupling reads now
\begin{equation}
\Delta_{\rm 2D}=\frac{1}{2}\left(\frac{m^*V}{\hbar^2}\right)^2 E_0,
\label{gap2dSO}
\end{equation}
where the bosonic energy $\omega_0$ is no longer present and the
relevant energy scale is provided by $E_0$. Note also the
quadratic dependence of the binding energy $\Delta$ with respect to the
coupling strength $V$, and the complete absence of an isotope effect 
for the phonon-mediator case.

{\em Superconducting critical temperature} -
The above discussion of the single Cooper pair problem will be now
a guide to the following investigation of the superconducting
transition for finite (low) densities in fully interacting
systems.
To this end we consider a Rashba-Holstein model where the SO coupled
electrons interact with dispersionless bosons with energy $\omega_0$
through an $s$-wave coupling with matrix element $g$ \cite{grimaldi06b}.
The superconducting properties, and in particular the critical temperature
$T_c$, are evaluated within the Eliashberg framework
properly generalized in the presence of SO coupling, with general values
of $E_F/E_0$, including the low density case just discussed.
Note that due to the lack of inversion symmetry, a different
superconducting phase with mixed even/odd order parameter and
singlet/triplet symmetry can in principle be included \cite{frigeri04}.
We neglect here for simplicity this issue and focus
only on the $s$-wave singlet channel.

\begin{figure}
\includegraphics[width=0.43\textwidth]{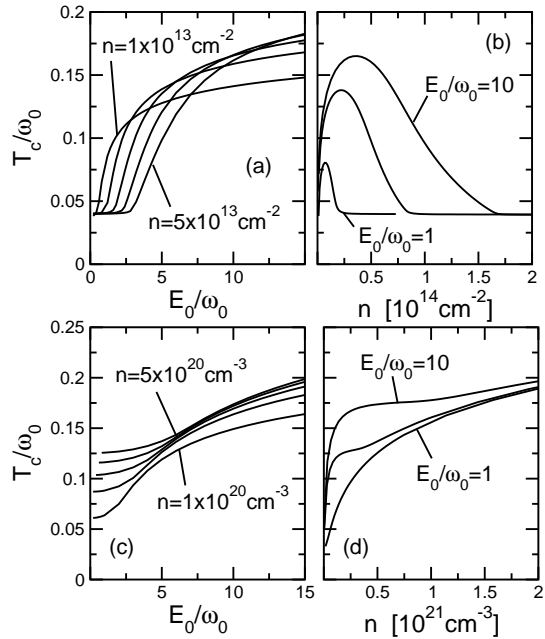}
\caption{ Panels (a) and (b): Superconducting critical
temperature $T_c$ as function of the Rashba spin-orbit energy
$E_0$ and of the electron density $n$, respectively, for the 2D
case. Panels (c) and (d): same quantities for the 3D system.}
\label{f-tc}
\end{figure}

In Fig.~\ref{f-tc}a,c we show the superconducting critical temperature
$T_c$ as a function of the SO Rashba energy $E_0$ for
different electron densities $n$.
In the figure the lower density values,
$n=10^{13}$ cm$^{-2}$ and  $n=10^{20}$ cm$^{-3}$,
correspond to Fermi energies $E_{\rm F}\simeq 24$ meV
and $E_{\rm F} \simeq 46$ meV
for the free electron gas in 2D and 3D respectively.
In addition, for a practical purpose one needs to introduce
a finite bandwidth cut-off $E_c$,
which is physically provided by the size of the Brillouin zone $k_c$.
We set $E_c=2000$($430$) meV which give $k_c \simeq 0.72$($0.33$) \AA$^{-1}$
respectively for the 2D and 3D case,
We get thus an energy-dimensional electronic DOS per unit cell
which, for the density values reported above in the absence
of SO coupling , is
$N_{\rm 2D}(E_{\rm F}=24 \mbox{meV})\simeq 5\cdot 10^{-4}$ meV$^{-1}$
and $N_{\rm 3D}(E_{\rm F}=46 \mbox{meV}) \simeq 12\cdot 10^{-4}$ meV$^{-1}$.
For all cases $\omega_0$ has been fixed at $\omega_0=20$ meV
and $g=5 \omega_0$.
With these values we obtain dimensionless coupling constants
$\lambda=2g^2 N(E_{\rm F})/\omega_0$ respectively
$\lambda_{\rm 2D}(E_{\rm F}=24 \mbox{meV})\simeq 0.5$ and 
$\lambda_{\rm 3D}(E_{\rm F}=46 \mbox{meV})\simeq 0.6$.
Fig.~\ref{f-tc}a,c shows a significant increase of $T_c$
as a function of $E_0$, in particular for low densities
where a small $E_0$ is sufficient to enter into the
$E_{\rm F} \lesssim E_0$ regime.
This holds true for both 2D and 3D systems, and the enhancement of $T_c$
can be as high as 300 \% with respect to the $E_0 \rightarrow 0$ limit.
Also interesting is the study of the superconducting critical temperature
as a function of the electron density $n$, as reported
in Fig. \ref{f-tc}b,d.
In these panels it is evident how
the $T_c$ vs. $n$ behavior reflects the effective dimensionality reduction
of the underlying non interacting DOS.
For example for the 2D case $T_c$ is nearly constant as a function of
electron density for $E_0=0$; this is consistent with the constant 
electron DOS.
In contrast, for finite $E_0$ the presence of strong peaks
in $T_c$ vs. $n$ reflects the 1D-like singularity
of the DOS. Note however that the retarded electron-boson interaction
gives rise to dynamical one-particle renormalization effects, automatically
taken into account in the Eliashberg equations, which smear 
the singularity of the bare DOS.

Similarly, in the $E_0 \rightarrow 0$ limit of the 3D case $T_c$
drops as the density $n$ is reduced due to vanishing
of $N_{\rm 3D}(E_{\rm F})\propto \sqrt{E_{\rm F}}$.
On the other hand the 2D character of the DOS with $E_0 \neq 0$
gives rise to an almost flat dependence of $T_c$ for sufficiently
low $n$,
with a critical temperature tuned by the Rashba energy $E_0$.
Both the 2D and 3D case thus show that the lack of
inversion symmetry not only affects the 
character of the order parameter, as discussed in several works
\cite{samokhin04,gorkov01,frigeri04},
but in principle can also lead to a substantial enhancement
of the superconducting pairing in the low density regime.
As explained above, this phenomenon is triggered by the topological 
change of the Fermi surface due to the strong SO interaction.

Let us discuss now the relevance
of our results in the context of real materials.
Concerning the 2D case, for instance,
surface states and low dimensional heterostructures
could be natural candidates for the search
of enhanced superconductivity.
In particular the issue of surface superconductivity \cite{gorkov06} has
recently been brought to attention due to its relevance for systems
like alkali-doped WO$_3$ \cite{wo3} where, for sufficiently low 
concentrations
of the alkali atoms, evidence of superconductivity confined
to the surface has been provided; this is precisely where strong SO coupling
arises from the confinement of the surface potential.
Interesting perspectives are also given by the
non-centrosymmetric superconductors, such as
CePt$_3$Si, Li$_2$Pd$_3$B, and Li$_2$Pt$_3$B, where
a strong Rashba energy arises from the lack of
inversion symmetry in the bulk crystal.
In the Li$_2$Pd$_3$B and Li$_2$Pt$_3$B compounds, in particular,
a large  SO coupling is accompanied by a strong electron-phonon
interaction \cite{pickett05}. In this case, of course, as in the heavy fermion
CePt$_3$Si case, a proper generalization of the present results
to the case of non parabolic bands is needed.

In summary, we have examined the impact of a strong Rashba 
spin-orbit interaction 
on superconductivity in a weakly coupled electron-boson system. The primary
result is an effective reduction in dimensionality of the electronic density
of states. This accomplishes what the Fermi sea did for the Cooper 
pair problem
in 3 dimensions --- it increased phase space at the Fermi level significantly
to allow binding for arbitrarily weak interactions. We then performed full
Eliashberg calculations of the critical temperature for a wide 
range of parameters;
these illustrate that the enhanced density of states has a 
significant impact
on the superconducting critical temperature. We suggest a search for higher
critical temperatures in materials with large spin-orbit coupling. In systems
where electron density can be varied one should be able to test some of the
trends reported here.

\begin{acknowledgments}

This work was supported in part by the
Natural Sciences and Engineering Research Council of Canada (NSERC),
by ICORE (Alberta), by the Canadian Institute for Advanced Research
(CIAR), and by the University of Geneva.

\end{acknowledgments}

\end{document}